\documentclass[aps,prl,twocolumn,showpacs,preprintnumbers,amsmath,amssymb]{revtex4}
\usepackage{epsfig}
\input epsf
\epsfclipon
\usepackage{epstopdf}

\begin{document}
\title{A minimal exactly solved model with the extreme Thouless effect}
\author{Agata Fronczak, Piotr Fronczak and Andrzej Krawiecki}
\affiliation{Faculty of Physics, Warsaw University of Technology,
Koszykowa 75, PL-00-662 Warsaw, Poland}
\date{\today}

\begin{abstract}
We present and analyze a minimal exactly solved model that exhibits a mixed-order phase transition known in the literature as the Thouless effect. Such hybrid transitions do not fit into the modest classification of thermodynamic transitions and as such, they used to be overlooked or incorrectly identified in the past. The recent series of observations of
such transitions in many diverse systems suggest that a new taxonomy of phase transitions is needed. The spin model we present due to its simplicity and possible experimental designs could bring us to this goal. We find the Hamiltonian of the model from which partition function is easily calculated. Thermodynamic properties of the model, i.e.
discontinuous magnetization and diverging susceptibility, are discussed. Finally, its generalizations and further research directions are proposed.   \end{abstract} 
\maketitle


Mixed-order phase transitions (MOT) are recently one of the most discussed and controversial topics in statistical physics and physics of complex systems. The transitions are also referred to as hybrid, because they combine features of continuous and discontinuous phase transitions.

A few years ago, the concept of hybrid transitions has become widely known through the phenomenon of explosive percolation, which at first was hailed as a discontinuous transition with critical fluctuations \cite{percolScience2009, percolPRL2010a}, but then it turned out to be either continuous or discontinuous (but not mixed-order) depending on various global rules and the dimensionality of the system \cite{percolPRL2010b, percolScience2011, percolNatCommun2012, percolScience2013a, percolScience2013b}. And although, in relation to the concept of hybrid transitions, the story mentioned was not successful it undoubtedly brought a very positive effect: It has renewed interest in various \emph{strange} phase transitions that are not covered by the contemporary theory and the classification scheme of these phenomena (see eg. \cite{JChPhysDNA1966, JChPhysFisher1966, PRLKafri2000, EPLLiu2012, PREBassler2015, JStatMechBassler2015, PRThouless1969, CMathPhysDyson1971, PRLBar2014, JStatMechBar2014}).

Why mixed-order transitions are so little known today? It seems that the main reason was the lack of simple and exactly solved physical models showing transitions of this kind. For example, in relation to continuous transitions, a belief is that \emph{"the general theory of critical phenomena and scaling theory (...) is a generalization of results first obtained for the Ising model"} \footnote{The statement is repeated after Ref.~\cite{McCoybook2010}, p.~277.}. This is probably due to the Onsager$'$s exact solution of the square lattice zero-field Ising model, that almost every physicist knows what the Ising model is and that one of the largest, still unresolved, challenges of statistical physics is the exact solution of this model in a non-zero-magnetic field. It is also a common knowledge that the one-dimensional Ising model with nearest-neighbor interactions has no phase transition at any positive temperature.

On the other hand, not everyone knows somewhat less trivial fact that the one-dimensional Ising model with a ferromagnetic coupling which decays as $1/r^\alpha$ with $1<\alpha<2$ has a standard phase diagram consisting of a single line of discontinuous transition at zero magnetic field, which ends at a critical point, where continuous transition occurs \cite{CMathPhysRuelle1968, CMathPhysDyson1969}. Furthermore, in the $\alpha=2$ case, instead of the continuous transition at the end of the first-order transition line, one can observe, what we call, mixed-order transition, in which a discontinuous jump in magnetization (first-order-like) accompanies diverging correlation length (second-order-like). Properties of this unusual transition in the $\alpha=2$ case were first suggested in 1969 by Thouless \cite{PRThouless1969}, thus getting the name of the \emph{Thouless effect}. And although many years passed since the Thouless effect in the prominent one-dimensional inverse distance squared Ising model (IDSI) was rigorously proved \cite{JPhysACardy1981, CMathPhysFrohlich1982, JStatPhysAizenman1988} the complete solution of the model is still lacking.

More recently, after long years of the only occasional research on the Thouless effect and hybrid phase transitions, a significant progress has been made in this area: A novel one-dimensional Ising model displaying spontaneous symmetry breaking discontinuous transition with diverging correlation length has been introduced and analysed \cite{PRLBar2014, JStatMechBar2014}, thus providing a new reference point for theoretical and numerical investigations of MOT. The new model mentioned is a truncated version of the prominent IDSI model. The purpose of this letter is to exhibit a completely new spin model for which the extreme Thouless effect can be rigorously demonstrated. The most important feature that distinguishes our model from the truncated IDSI is its extraordinary simplicity, which refers both to the very definition of the model, as well as the direct method of its complete analytical solution.

The model is defined as follows: We study $N$ distinguishable and noninteracting spins, which can have two states, $s_i=\pm 1$. With a probability $q$, we randomly select one of positive spins ($+1$) and change its state to the opposite ($-1$). With a probability $1-q$ we perform the reverse action, i.e. a randomly selected negative spin is flipped 

The model specified is clearly ergodic, therefore there must exist its equilibrium representation in the sense of the canonical ensemble. Thus, assuming that the probability of the system to be found in a certain spin configuration, $\Omega$, is given by the standard canonical distribution, $P(\Omega)\sim e^{-\mathcal{H}(\Omega)}$, the hamiltonian of the model, $\mathcal{H}(\Omega)$, can be recursively recovered from the detailed balance condition, which underlies the given dynamical definition. In the following, the hamiltonian reconstruction procedure is briefly discussed. Then, with the hamiltonian, the partition function of the model is exactly calculated. Finally, thermodynamic properties of the hybrid transition in the model are demonstrated, which rely on the discontinuously changing average magnetization per spin and the diverging magnetic susceptibility.

First, given that $\Omega$ and $\Omega'$ are two spin configurations, which differ from each other by the state of only one spin (let us say $s_i$, in such a way that $s_i(\Omega)=+1$ and $s_i(\Omega')=-1$) the transition probabilities among these configurations are equal to:
\begin{equation}\label{pO1O2}
p(\Omega\rightarrow\Omega')=\frac{q}{N_{+}(\Omega)},
\end{equation}
and
\begin{equation}\label{pO2O1}
p(\Omega'\rightarrow\Omega)=\frac{1-q}{N_{-}(\Omega')},
\end{equation}
where $N_{+}(\Omega)$ and $N_{-}(\Omega')$ stand for the number of positive and negative spins in the configurations specified, and
\begin{equation}\label{NpNm}
N_{-}(\Omega')=N-N_{+}(\Omega')=N+1-N_{+}(\Omega).
\end{equation}
Then, inserting Eqs.~(\ref{pO1O2})$-$(\ref{NpNm}) into the well-known expression for the detailed balance condition,
\begin{equation}\label{DB}
\frac{p(\Omega\rightarrow\Omega')}{p(\Omega'\rightarrow\Omega)}=\frac{P(\Omega')}{P(\Omega)} =e^{\mathcal{H}(\Omega)-\mathcal{H}(\Omega')},
\end{equation}
one gets the following recurrence relation for the hamiltonian of the model studied:
\begin{equation}\label{Hrec}
\mathcal{H}(\Omega)=\mathcal{H}(\Omega')+\ln\left(\frac{q}{1-q}\right)+ \ln\left(\frac{N_{-}(\Omega)+1}{N_{+}(\Omega)}\right).
\end{equation}

Eq.~(\ref{Hrec}) can be solved using standard methods (see \cite{bookConcreteMath}). It can also be solved \emph{step by step}, by imagining that we successively flip all the remaining positive spins. The flipping procedure ends when the system reaches the ground state, $\Omega_0$, when all the spins are negative. The solution of Eq.~(\ref{Hrec}) is as follows:
\begin{equation}\label{H0}
\mathcal{H}(\Omega)=\mathcal{H}(\Omega_0)+N_{+}(\Omega)\ln\left(\frac{q}{1-q}\right) +\ln{N\choose N_{+}(\Omega)}.
\end{equation}

Given the hamiltonian of the model, we now proceed to calculate its partition function, $Z(q)$, which is a normalization constant for the canonical distribution. It is obtained by summing Boltzmann factors, $e^{-\mathcal{H}(\Omega)}$, for all possible spin configurations, and it turns out to be mere sum of geometric series:
\begin{eqnarray}\label{Z0}
Z(q)&=&\sum_{\Omega}{N\choose N_+(\Omega)}^{-1}\left(\frac{1-q}{q}\right)^{N_+(\Omega)}\\
\label{Z1}&=&\sum_{N_+=0}^N a^{N_+}=\frac{1-a^{N+1}}{1-a},
\end{eqnarray}
where
\begin{equation}\label{defa}
a=\frac{1-q}{q}.
\end{equation}

\begin{figure}
\centerline{\epsfig{file=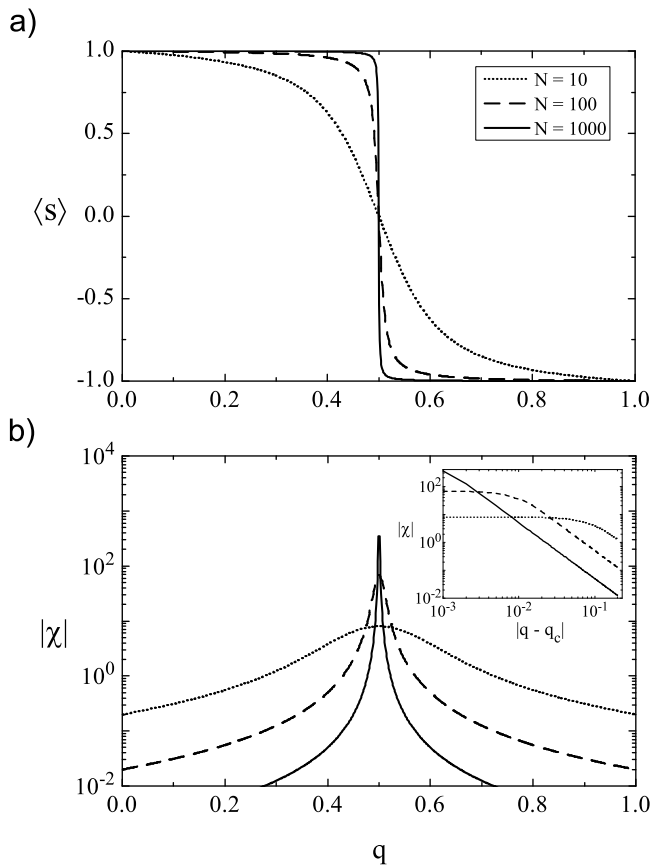,width=\columnwidth}}
\caption{Thermodynamic properties of the spin model for various system sizes $N$. a) Average magnetization per spin, $\langle s\rangle$, vs. the parameter $q$ as given by Eq.~(\ref{means0b}). b) Magnetic susceptibility $\chi$ as defined by Eq.~(\ref{chi1}). Inset: Scaling behaviour of $\chi$ vs. the distance from the critical point, $|q-q_c|$.}
\label{fig1}
\end{figure}

Now, thermodynamic properties of the model can be easily obtained. In particular, the average number of positive spins is simply given by
\begin{equation}\label{srNp}
\langle N_+\rangle=a\frac{\partial\ln Z}{\partial a},
\end{equation}
from which the average magnetization per spin can be derived
\begin{eqnarray}\label{means0a}
\langle s\rangle &=&\frac{1}{N}\left\langle\sum_{i=1}^Ns_i\right\rangle=\frac{\langle N_+-N_-\rangle}{N}\\\label{means0b} &=&1-\frac{2}{(a-1)N}+\frac{2(N+1)}{(a^{N+1}-1)N},
\end{eqnarray}
which shows discontinuity in the thermodynamic limit:
\begin{equation}\label{means1}
\lim_{N\rightarrow\infty}\langle s\rangle=\left\{ \begin{array}{lcl}
+1 & \mbox{for } & q<q_c \\
\;\;\;0 & \mbox{for } & q=q_c\\
-1 & \mbox{for } & q>q_c
\end{array}\right.,
\end{equation}
where
\begin{equation}\label{qc}
q_c=\frac{1}{2}.
\end{equation}
Furthermore, when differentiating Eq.~(\ref{means0b}) with respect to $q$, one can see that the extreme Thouless effect, Eq.~(\ref{means1}), is accompanied by the critical-like divergence of the magnetic susceptibility
\begin{equation}\label{chi1}
\chi=\frac{\partial \langle s\rangle}{\partial q} \sim\frac{1}{N}\left|q-q_c\right|^{-2},
\end{equation}
thus providing the evidence of the hybrid transition in the spin model introduced (see Fig.~\ref{fig1}).

\begin{figure}
\centerline{\epsfig{file=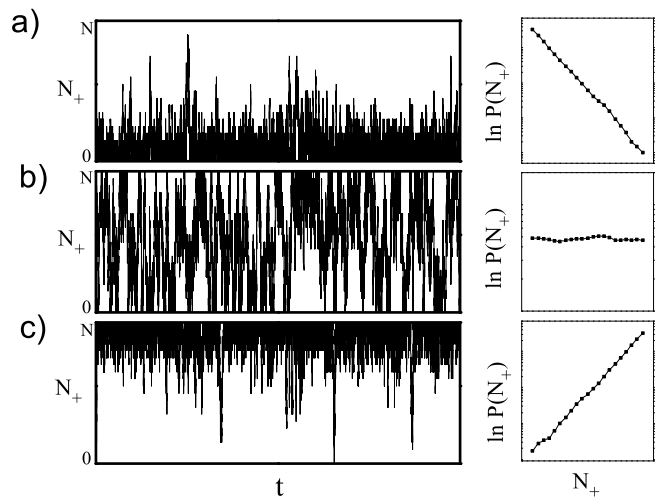,width=\columnwidth}}
\caption{Time traces of $N_+$ for three different cases of the biased random walk which is confined by reflecting walls to the region $0\leq N_+\leq N$, where $N=20$: a) $q=0.6$, b) $q=0.5$, c) $q=0.4$. The corresponding histograms of $N_+$ are shown on the right site of time plots.}
\label{fig2}
\end{figure}

At first glance, it is hard to intuitively understand, how the extreme Thouless effect arises in the system studied. To facilitate understanding of the phenomenon, the concept of the $q$-biased one-dimensional discrete-space and -time random walker, whose position $N_+(t)$ is confined by reflecting walls to the region $0\leq N_+\leq N$, proves useful. The $q$-bias means that $N_+$ decreases by one with probability $q$ and increases by one with probability $1-q$, completely the same as in the spin model (see Fig.~\ref{fig2}).

For $q=\frac{1}{2}$ the random walk is symmetric and its stationary occupation probability distribution is uniform \cite{RW1}. The uniform distribution also arises from the canonical approach to the spin model:
\begin{eqnarray}\label{PNp0}
P(N_+)&=&\sum_{\Omega^*}P(\Omega)=\frac{a^{N_+}}{Z(a)}\\
\label{PNp1}&\stackrel{a=1}{=}&\frac{1}{N+1},
\end{eqnarray}
where the summation extends over all spin configurations, $\Omega^*$, such that the number of positive spins is fixed and equal to $N_+$. Correspondingly, in this case, the average number of positive spins is half of the number of all spins, $\langle N_+\rangle=\frac{N}{2}$, which amounts to $\langle s\rangle=0$, and fluctuations of $N_+$ spread all over the allowed range of variation of the parameter.

On the other hand, if $q\neq\frac{1}{2}$ the random walk is asymmetric. One drift direction is preferred over the other and the stationary occupation probability distribution becomes skewed. In particular, for $q>\frac{1}{2}$ (i.e. for $a<1$), in systems large enough, the probability distribution, Eq.~(\ref{PNp1}), simplifies to the geometric distribution:
\begin{equation}\label{PNp2}
P(N_+)\stackrel{N\gg 1}{=}(1-a)a^{N_+}.
\end{equation}
This causes that, in the thermodynamic limit, the average number of positive spins is finite, $\langle N_+\rangle=a/(1-a)$, and therefore small compared to the system size. As a result, relative fluctuations of $N_+$ disappear. The same happens with the magnetic susceptibility, cf. Eq.~(\ref{chi1}).
Obviously, the analogous behaviour is observed for $q<\frac{1}{2}$ (i.e. $a>1$). In this case, however, since $N_-=N-N_+$, one can show that Eq.~(\ref{PNp0}) can be simplified to geometric distribution for the number of negative spins:
\begin{equation}\label{PNm3}
P(N_-)\stackrel{N\gg 1}{=}\left(1-\frac{1}{a}\right) \left(\frac{1}{a}\right)^{N_-}.
\end{equation}
Other properties of the system for $q<\frac{1}{2}$ can be easily calculated, in the same way as for $q>\frac{1}{2}$.

To conclude, we have presented and strictly solved a novel spin model that displays the extreme Thouless effect, i.e. a discontinuous jump in magnetization (from $+1$ to $-1$), which is accompanied by the diverging susceptibility. The spin model introduced is important for a number of reasons. First, it is the simplest and so far the only exactly solved model with hybrid transition \footnote{The recently discussed truncated version of the one-dimensional inverse distance squared Ising model \cite{PRLBar2014,JStatMechBar2014}, which is the prototypical example of mixed-order phase transitions, has been solved in the so-called \emph{ideal gas of clusters} approximation.}. And although the one-parameter character of model can be considered its shortcoming, the possible extensions of the model (motivated by its dual relationship with both spin-lattice systems and random-walk processes) open a rich perspective for future research. In particular, since the current version of the model does not explicitly specify the interaction between spins, consideration of such effects could enrich the behaviour of the system by the appearance of continuous and discontinuous phase transitions. For example, it is a great challenge to understand general mechanisms leading to hybrid transitions. This challenge would be easier to meet if we could confirm that hybrid transitions always appear in the area between continuous and discontinuous transitions (which are already well understood), as it was observed in the prototypical truncated IDSI model \cite{PRLBar2014,JStatMechBar2014}.

Second, the model introduced can serve as a convenient testbed for new theoretical and numerical approaches (such as the generalized method of finite-size scaling) dedicated to the study of MOT. The controversial story of explosive percolation \cite{percolScience2009, percolPRL2010a, percolPRL2010b, percolScience2011, percolNatCommun2012, percolScience2013a, percolScience2013b} showed that the issue mentioned is really important, because distinguishing, in numerical studies of finite systems, a discontinuity from a sharp continuous transition is a very difficult task. Therefore, one can assume that in the near future the problem of numerical studies of hybrid transitions (which has been already signalled in e.g.~\cite{PRLLuijten2001}) will be a major challenge for computational statistical physics.

We thank T.~Ryczkowski and A.M.~Chmiel for helpful discussions. The support from the National Science Centre in Poland (grant no. 2012/05/E/ST2/02300) is gratefully acknowledged.

\end{document}